\documentclass[sn-mathphys,iicol,Numbered]{sn-jnl}
\usepackage{graphicx}%
\usepackage{multirow}%
\usepackage{amsmath,amssymb,amsfonts}%
\usepackage{amsthm}%
\usepackage{mathrsfs}%
\usepackage[title]{appendix}%
\usepackage{xcolor}%
\usepackage{textcomp}%
\usepackage{manyfoot}%
\usepackage{booktabs}%
\usepackage{algorithm}%
\usepackage{algorithmicx}%
\usepackage{algpseudocode}%
\usepackage{listings}%
 
\theoremstyle{thmstyleone}%
\theoremstyle{thmstyletwo}%

\theoremstyle{thmstylethree}%

\usepackage[normalem]{ulem}  

\renewcommand{\sout}{\bgroup \color{red} \ULdepth=-.5ex \ULset}

\newcommand{\hyt}{$^3_\Lambda \text{H}$}
\newcommand{\lam}{$\Lambda$}

\usepackage{tikz,xcolor,hyperref}
\definecolor{lime}{HTML}{A6CE39}
\DeclareRobustCommand{\orcidicon}{
	\begin{tikzpicture}
	\draw[lime, fill=lime] (0,0) 
	circle [radius=0.16] 
	node[white] {{\fontfamily{qag}\selectfont \tiny ID}};
	\draw[white, fill=white] (-0.0625,0.095) 
	circle [radius=0.007];
	\end{tikzpicture}
	\hspace{-2mm}
}

\foreach \x in {A, ..., Z}{%
	\expandafter\xdef\csname orcid\x\endcsname{\noexpand\href{https://orcid.org/\csname orcidauthor\x\endcsname}{\noexpand\orcidicon}}
}

\newcommand\x{\xi}

\begin{document}

\title[Article Title]{From Hyperons to Hypernuclei: A New Route to Unravel Proton Spin Polarization}

\author[1,2]{\fnm{Dai-Neng} \sur{Liu}\orcidF{}}  \email{dnliu17@fudan.edu.cn}
\author[1,2]{\fnm{Yun-Peng} \sur{Zheng}\orcidG{}}
\email{ypzheng23@m.fudan.edu.cn}
\author[1,2,3]{\fnm{Wen-Hao}\sur{Zhou}\orcidH{}}
\email{zhouwenhao@xaau.edu.cn}
\author[1,2]{\fnm{Jin-Hui}  \sur{Chen}\orcidB{}}\email{chenjinhui@fudan.edu.cn}
\author[4]{\fnm{Che Ming}   \sur{Ko}  \orcidC{}}\email{ko@comp.tamu.edu}
\author*[1,2]{\fnm{Yu-Gang} \sur{Ma}  \orcidD{}}\email{mayugang@fudan.edu.cn}
\author*[1,2]{\fnm{Kai-Jia} \sur{Sun} \orcidA{}}\email{kjsun@fudan.edu.cn}
\author[1,2]{\fnm{Song}  \sur{Zhang}\orcidE{}}\email{song\_zhang@fudan.edu.cn}

\affil[1]{\orgdiv{Key Laboratory of Nuclear Physics and Ion-beam Application (MOE), Institute of Modern Physics}, \orgname{Fudan University}, \orgaddress{\city{Shanghai}, \postcode{200433}, \country{China}}}
\affil[2]{\orgdiv{Shanghai Research Center for Theoretical Nuclear Physics}, \orgname{NSFC and Fudan University}, \orgaddress{\city{Shanghai}, \postcode{200438}, \country{China}}}
\affil[3]{\orgdiv{Faculty of Science}, \orgname{Xihang University}, \orgaddress{\city{Xi’an}, \postcode{710077}, \country{China}}}
\affil[4]{\orgdiv{Cyclotron Institute and Department of Physics and Astronomy}, \orgname{Texas A\&M University}, \orgaddress{\city{College Station}, \postcode{TX 77843}, \country{USA}}}

\abstract{  Ultra-relativistic nuclear collisions create the   quark–gluon 
plasma (QGP) known as the hottest, least viscous, and most vortical fluid ever produced in terrestrial laboratories. Its vortical structure has been uncovered through the spin polarization of Lambda ($\Lambda$) hyperons, attributed to the spin–orbit coupling that transfers the system’s orbital angular momentum to the quark spin, which is then inherited by hadrons via quark recombination or coalescence. However, $\Lambda$ polarization reflects primarily the strange-quark component, leaving the spin dynamics of the up and down quarks largely unexplored. Although the proton is an ideal probe, its stability makes direct measurements experimentally challenging.  
Here, we propose to unravel proton spin polarization via hypertriton (\hyt) measurements, exploiting the fact that spin information is preserved when polarized nucleons and $\Lambda$  coalesce to form hypertriton. We show that, over a broad range of collision energies, the polarizations of proton, $\Lambda$, and hypertriton  are related by a simple linear scaling law. Since both $\Lambda$ and hypertriton  polarizations can be measured via their self-analyzing weak decays,  this linear relation provides a practical experimental avenue for accessing spin polarizations of protons and neutrons–the dominant baryonic degrees of freedom in nuclear collisions.
}




\maketitle

\begin{figure*}[!t]
\centering 
\includegraphics[width=16cm]{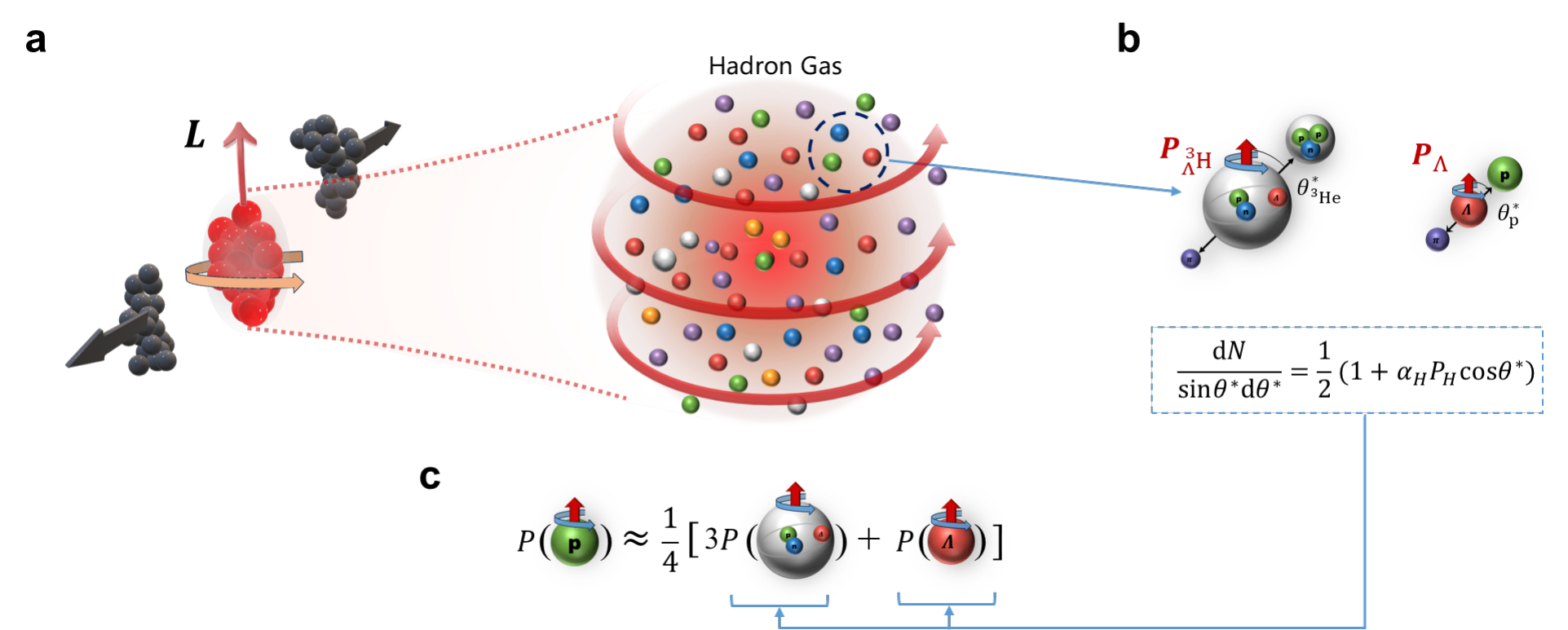} 
\caption{
\textbf{Schematic illustration of decoding proton spin polarization in heavy-ion collisions.}
\textbf{a}, In non-central heavy-ion collisions, the overlap region of colliding nuclei carries a large orbital angular momentum $L$, leading to spin polarization of quarks through the spin-orbit interaction. Hadrons emerging from the QGP at hadronization acquire spin polarization via coalescence of polarized quarks, and subsequently evolve through the hadronic phase until the kinetic freeze-out. \textbf{b}, Hypertritons (\hyt) are polarized when formed via the coalescence of polarized nucleons and $\Lambda$ hyperons at the kinetic freeze-out. Both \hyt~and $\Lambda$ undergo self-analyzing weak decays (e.g. $\Lambda\rightarrow \pi^-+p$ and \hyt$\rightarrow \pi^-+^3$He), allowing their spin polarizations to be determined from the angular distributions ($dN/\sin\theta^*d\theta^*$) of their decay products. \textbf{c}, The proton spin polarization ($\mathcal{P}_p$) can then be reconstructed from the polarizations of \hyt~and \lam~using a model-validated linear relation.
}
\label{pic:polarization} 
\end{figure*}

Quantum chromodynamics (QCD)—the fundamental theory of the strong interaction—predicts that high-energy nuclear collisions can create a novel state of matter composed of deconfined quarks and gluons, known as the quark–gluon plasma (QGP)~\cite{STAR:2005gfr}. This exotic phase evolves as  a nearly perfect fluid and is believed to have existed in the first few microseconds after the Big Bang.

In semi-central collisions, the QGP carries substantial orbital angular momentum, on the order of $10^3\hbar$, oriented perpendicular to the reaction plane~\cite{STAR:2017ckg}.   Through the spin–orbit or spin–vorticity coupling, part of this angular momentum can be transferred to the spin degrees of freedom of quarks, and subsequently inherited by hadrons via quark coalescence or recombination at hadronization~\cite{Liang:2004ph}.  This mechanism gives rise to  the phenomenon of global spin polarization~\cite{Liang:2004ph, Voloshin:2004ha,Becattini:2007sr,Becattini:2020ngo}, analogous to the Barnett and Einstein–de Haas effect in condensed matter~\cite{Barnett,einstein1915}. Experiments at the Relativistic Heavy Ion Collider (RHIC) have observed global spin polarization of $\Lambda$ and $\bar{\Lambda}$ hyperons, confirming that the QGP is the most vortical fluid ever observed~\cite{STAR:2017ckg}.  Since then, significant progress has been made to understand the role of spin degrees of freedom in the dynamics of QGP~\cite{Fu:2021pok,Becattini:2021iol,Sheng:2022wsy,STAR:2022fan, Chen:2023hnb,Becattini:2024uha, Niida:2024ntm,Chen:2024hki}. 

Despite these advances, a crucial piece in our understanding of spin dynamics in hot and dense QCD matter remains missing: the spin polarization of protons and neutrons—the fundamental constituents of visible matter and dominant baryonic degrees of freedom in nuclear collisions—has not been measured. Measuring proton spin polarization is of particular importance for addressing the polarization of light quarks in the QGP (beyond what $\Lambda$ polarization reveals about  strange quarks) and the spin-dependent transport phenomenon~\cite{Liu:2023nkm,Xu:2025uwd} such as the hadronic spin Hall effect~\cite{Liu:2020dxg}. Unlike hyperons, however, these baryons do not decay within the detector volume, making conventional polarization analysis via self-analyzing weak decays infeasible. A direct measurement would require external polarimeters based on secondary interactions, which has not yet been implemented in current heavy-ion detectors~\cite{Liang:2025owx}.

Here we propose a novel approach to access proton spin polarization by combining independent measurements of $\Lambda$ and $^{3}_{\Lambda}\mathrm{H}$ polarizations, both of which can be determined from their parity-violating weak decays, e.g., $\Lambda\rightarrow \pi^-+p$ and \hyt$\rightarrow \pi^-+^3$He. This approach is based on the idea that  when polarized protons, neutrons, and $\Lambda$ hyperons coalesce into \hyt~(a loosely bound state of a spin-1 deuteron and a spin-1/2 $\Lambda$ hyperon~\cite{ALICE:2022sco}) at the kinetic freeze-out in nuclear collisions, their spin information is preserved~\cite{Liu:2023nkm, Sun:2025oib}.  Based on comprehensive model simulations for  non-central Au+Au collisions over a broad range of collision energies ($\sqrt{s_{NN}} = 7.7$–200 GeV), including spin-dependent hadronic interactions and resonance decays, we show that proton spin polarization ($\mathcal{P}_p$) can be reliably reconstructed via a simple linear relation:
\begin{eqnarray}
\mathcal{P}_p \approx \frac{1}{4}\left(3\mathcal{P}_{^3_\Lambda\mathrm{H}} + \mathcal{P}_\Lambda\right). \label{eq:NucleonPolarization}
\end{eqnarray}
This relation is a direct reflection of the fact that in the spin wavefunction of the \hyt~($J^\pi=\frac{1}{2}^+$~\cite{STAR:2017gxa}), the $\Lambda$ spin aligns with the total spin with 1/3 probability and anti-aligns with 2/3 probability. A schematic illustration of the proposed method for decoding proton spin polarization via $\Lambda$ and  \hyt~production is shown in Fig.~\ref{pic:polarization}.

\begin{figure*}[!t]
    \centering
    \includegraphics[scale=0.56]{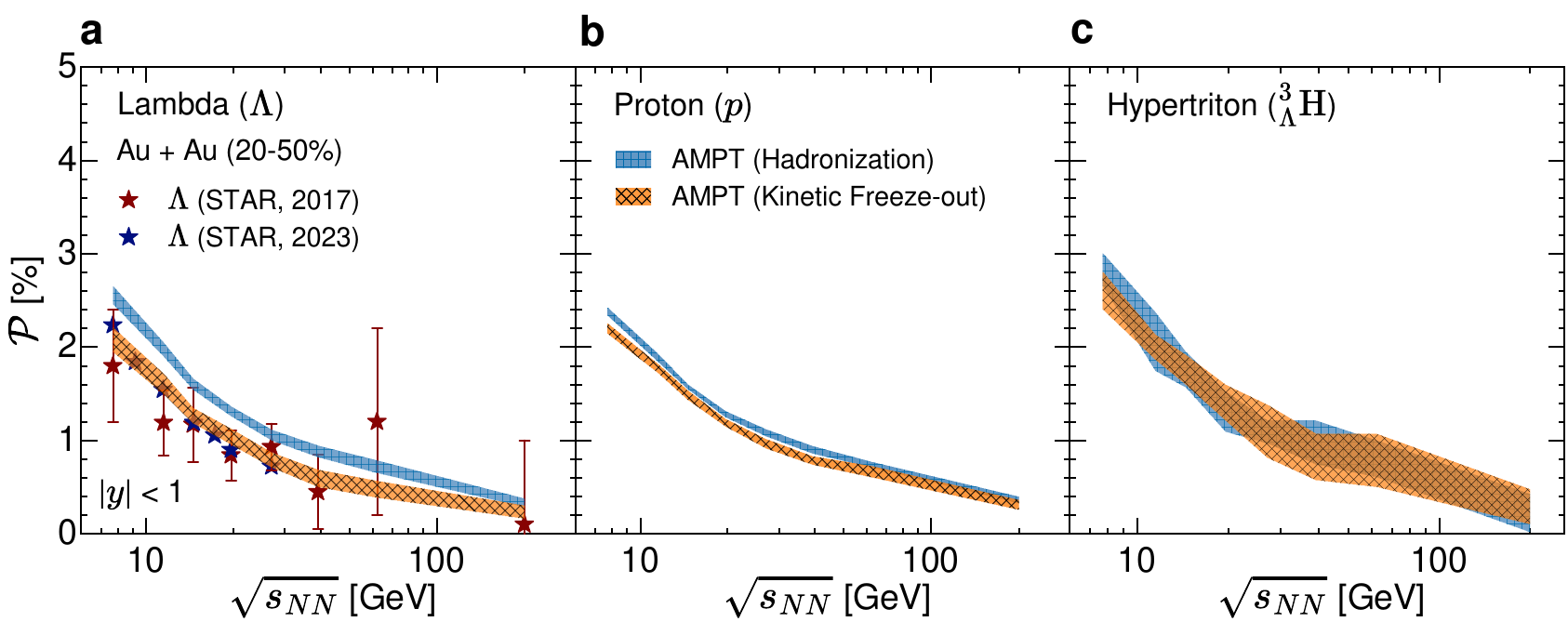} 
    \caption{\textbf{Collision energy dependence of global spin polarizations of $\Lambda$ hyperons, protons, and hypertritons.} 
    \textbf{a}, Global polarization of mid-rapidity ($|y|<1$) $\Lambda$ hyperons in Au+Au collisions at $\sqrt{s_{NN}} = 7.7$–200 GeV  and 20–50\% centrality. The blue and orange bands represent model predictions from  an extended AMPT  model at hadronization and kinetic freeze-out stages, respectively, using an impact parameter $b = 8$ fm. Experimental data for $\Lambda$ hyperons from the STAR collaboration are shown with error bars, where the uncertainties combine statistical and systematic contributions~\cite{STAR:2017ckg, STAR:2023nvo}. \textbf{b}, Corresponding global polarization of proton at hadronization (blue band) and kinetic freeze-out (orange band). \textbf{c}, Same as in \textbf{b}, but for hypertritons.  Uncertainties  in theoretical results   are from statistics.}
    \label{fig:global_L}
\end{figure*}

\bmhead{Global spin polarizations of $\Lambda$ hyperon and proton in heavy-ion collisions}
We employ a multi-phase transport (AMPT) model~\cite{Lin:2004en} to simulate the evolution of heavy-ion collisions, which proceeds through four main stages: initialization, partonic evolution, hadronization via quark coalescence, and the subsequent hadronic rescatterings and decays. To study spin dynamics of hadrons, we extend the AMPT model to incorporate spin evolution during the hadronic phase—from hadronization to kinetic freeze-out. Particles with non-zero spins, such as nucleons and $\Lambda$ hyperons, acquire polarization at hadronization through spin–vorticity coupling, which is computed using the Cooper–Frye formula~\cite{Cooper:1974mv,Becattini:2013fla}. During the hadronic phase, their spin polarizations are further modified by spin-dependent processes, including resonance decays and scatterings that can induce spin flips. After kinetic freeze-out, where interactions cease, the spin states of particles remain unchanged and spin polarization can be determined. Using this extended framework (see ``Methods'' section), we first investigate the global spin polarizations of nucleons and $\Lambda$ hyperons in non-central Au+Au collisions at $\sqrt{s_{NN}} = 7.7$–200 GeV. 

Figure \ref{fig:global_L} (a) shows the energy dependence of the global  polarization of mid-rapidity ($|y|<1$) \lam~hyperons along the direction of total angular momentum at hadronization (blue band) and kinetic freeze-out stages (orange band)  in collisions at impact parameter $b=8$ fm, corresponding to 20-50\% centrality.  For comparison, experimental data from the STAR Collaboration are represented by  filled symbols for $\Lambda$ hyperons~\cite{STAR:2017ckg,  STAR:2023nvo} with combined statistical and systematic uncertainties. 
The global spin polarization of $\Lambda$ hyperons decreases with increasing beam energies,  in good quantitative agreement with the STAR data as well as other theoretical calculations summarized in Refs.~\cite{Huang:2020xyr, Becattini:2020ngo, Chen:2024aom}.  Note that the experimental data points are rescaled using the updated decay parameter of $\Lambda$ hyperon  $\alpha_\Lambda=0.732 \pm 0.014$~\cite{ParticleDataGroup:2024cfk}.

Our results show that the $\Lambda$ polarization at kinetic freeze-out is reduced by approximately 20\% relative to its value at hadronization. This suppression arises from combined effects of resonance feed-down and spin-flip scatterings during the hadronic phase. Specifically, the feed-down of primary $\Sigma^*$ resonances tends to enhance the $\Lambda$ polarization~\cite{Li:2021zwq}, while processes such as $\pi\Lambda \rightarrow \Sigma^*(1385) \rightarrow \pi\Lambda$ scatterings and $\Sigma^0 \rightarrow \gamma \Lambda$ decays lead to a reduction. The obtained magnitude of suppression from hadronic re-scatterings and decays is consistent with recent studies using rate equations~\cite{Sung:2024vyc,Sung:2025pdj}.

Figure~\ref{fig:global_L} (b) depicts the beam energy dependence of proton spin polarization at the hadronization stage (blue band) and at kinetic freeze-out (orange band). At hadronization, the spin polarizations of nucleons and $\Lambda$ hyperons are found to be close to each other, owing to their similar masses and the same spin quantum numbers. Unlike the case for $\Lambda$ hyperons, whose polarization is significantly reduced during the hadronic phase, the nucleon polarization remains almost unchanged between hadronization and kinetic freeze-out stages. This difference results from a near cancellation between two competing effects on proton spin polarization: an enhancement from feed-down decays of primary $\Delta(1232)$ resonances (with spin-3/2), and a suppression from spin depolarization induced by quasi-elastic scatterings with pions, particularly through the process $\pi N \rightarrow \Delta \rightarrow \pi N$. 

According to the quark recombination or coalescence model, the $\Lambda$ hyperon's polarization ($\mathcal{P}_\Lambda$) directly reflects that of the strange quark ($\mathcal{P}_s$), as the up-down ($ud$) diquark inside the $\Lambda$ forms a spin singlet and contributes no net spin, i.e., $\mathcal{P}_\Lambda \approx \mathcal{P}_s$~\cite{Liang:2004ph,Liang:2004xn}.   Thus, the measured $\Lambda$ polarization provides little information on the spin polarization of $u$ and $d$ quarks in the QGP. In contrast, the proton spin polarization ($\mathcal{P}_p$) is directly related to the spin polarizations of $u$ and $d$ quarks and can be approximated in the SU(6) quark model as $\mathcal{P}_p \approx (4\mathcal{P}_u - \mathcal{P}_d)/3$ ~\cite{Liang:2004ph}. Therefore, the experimental measurement of proton spin polarization would provide valuable insights into the spin dynamics of light quarks in the QGP. Moreover, it would also shed light on the long-standing proton spin puzzle~\cite{Ji:2020ena}, given that first-principles Lattice QCD calculations indicate that quarks account for only about 38\% of the proton’s intrinsic spin~\cite{Alexandrou:2020sml}. 

 \bmhead{Global spin polarization of hypertriton} 
 We next investigate the spin polarization of the \hyt, assuming that it is formed by coalescence of freeze-out nucleons and $\Lambda$ hyperons. In this case, the nucleon polarization is carried over and encoded in the polarization of the resulting \hyt. The \hyt~can be viewed as a loosely bound state of a spin-1 deuteron and a spin-1/2 $\Lambda$ with total spin-parity $J^\pi = \frac{1}{2}^+$. This is supported by several experimental observations including its comparable lifetime with free $\Lambda$~\cite{STAR:2021orx,ALICE:2022sco}, extremely small $\Lambda$ separation energy of $B_\Lambda \approx 0.164$ MeV~\cite{STAR:2019wjm, Chen:2023mel, HypernuclearDataBase}, the decay branching ratio~\cite{ STAR:2017gxa}, as well as recent calculations based on chiral hyperon–nucleon and hyperon–nucleon–nucleon interactions~\cite{Le:2024rkd}. The possibility of other spin configurations is less likely but has recently  been  explored in Ref.~\cite{Sun:2025oib}.

\begin{figure*}[!t]
    \centering
    \includegraphics[scale=0.6]{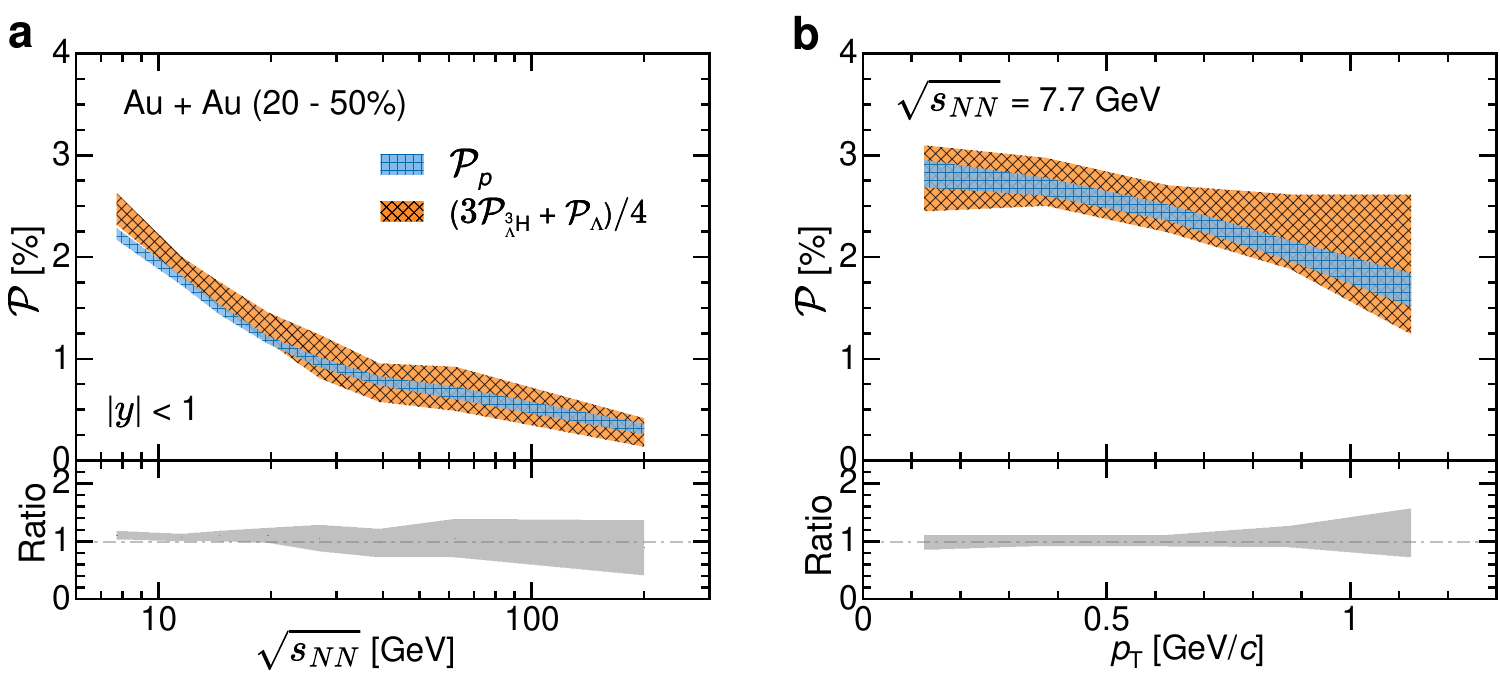}
    \caption{\textbf{Decoding proton spin polarization in heavy-ion collisions.} 
    \textbf{a}, Collision-energy dependence of global proton spin polarization in non-central Au+Au collisions  at $\sqrt{s_{NN}} = 7.7$–200 GeV, evaluated at mid-rapidity ($|y|<1$). \textbf{b}, Transverse-momentum dependence of proton spin polarization at the energy of 7.7 GeV. The lower panels show the ratio of the extracted value of proton polarization (orange bands) to its true value (blue bands) in model calculations. Coloured bands represent statistical uncertainties in theoretical results. }
    \label{fig:global_nucleon}
\end{figure*}

The coalescence model based on the density matrix formalism~\cite{Scheibl:1998tk} has been shown to successfully describe \hyt~production in high-energy nuclear collisions~\cite{Sun:2018mqq, Bellini:2018epz, ALargeIonColliderExperiment:2021puh}.
The  number of \hyt~ produced in the coalescence model is given by~\cite{Sun:2025oib} (see ``Methods" for more details)
\begin{eqnarray}
\label{eq:polaCoal}
  N_{^3_\Lambda \text{H},\pm\frac{1}{2}} &=&\sum_{np\Lambda} g_{\text{\hyt},\pm \frac{1}{2}}  W_{^3_\Lambda \text{H}}({\bf x}_n,{\bf x}_p,{\bf x}_\Lambda;{\bf p}_n,{\bf p}_p,{\bf p}_\Lambda),\notag \\ 
\end{eqnarray}
where the summation is over all possible combinations of neutron, proton, and $\Lambda$ hyperon;  and $W_{^3_\Lambda \text{H}}$ is the Wigner function representing the internal structure of the \hyt~in phase space. The  $ g_{\text{\hyt},\pm \frac{1}{2}}$ in the above equation denotes the spin statistical factor,
\begin{eqnarray}
g_{\text{\hyt},\pm \frac{1}{2}}&=& \frac{2}{3} w_{n,\pm \frac{1}{2}} w_{p,\pm \frac{1}{2}} w_{\Lambda \mp \frac{1}{2}} \notag\\&+&\frac{1}{6} w_{n,\mp \frac{1}{2}} w_{p,\pm \frac{1}{2}} w_{\Lambda \pm \frac{1}{2}}\notag \\
  &+&\frac{1}{6} w_{n,\pm \frac{1}{2}} w_{p,\mp \frac{1}{2}} w_{\Lambda \pm \frac{1}{2}}, 
\end{eqnarray}
where $w_{i, \pm \frac{1}{2}} = \frac{1}{2}(1 \pm \mathcal{P}_i)$ are the diagonal elements of the spin density matrix, with $\mathcal{P}_i$ denoting the spin polarization of particle $i$ (nucleon or $\Lambda$ hyperon).  For the case of unpolarized nucleons and $\Lambda$ hyperons (i.e., $\mathcal{P}_i = 0$), the statistical factor simplifies to  $g_{\text{\hyt},+\frac{1}{2}}=g_{\text{\hyt},- \frac{1}{2}} =1/8$. In terms of the numbers of \hyt~in spin up and spin down states,  the \hyt~spin polarization can be determined by $\mathcal{P}_{^3_\Lambda\text{H}}=(N_{^3_\Lambda \text{H},\frac{1}{2}}-N_{^3_\Lambda \text{H},-\frac{1}{2}})/(N_{^3_\Lambda \text{H},\frac{1}{2}}+N_{^3_\Lambda \text{H},-\frac{1}{2}})$.

Figure \ref{fig:global_L} (c) shows the collision energy dependence of  \hyt~polarization, which also decreases with increasing beam energies,  resembling the trends for the polarizations of protons and $\Lambda$ hyperons.  The polarization of \hyt~at the kinetic freeze-out stage is almost identical to that at the hadronization stage. Besides, the polarization of \hyt~is slightly larger than that of $\Lambda$ hyperon, which is advantageous for experimental measurements.

As mentioned earlier, both of $\mathcal{P}_\Lambda$ and $\mathcal{P}_{^3_\Lambda \text{H}}$ can be determined from their parity-violating weak decays, e.g., $\Lambda\rightarrow \pi^-+p$ and \hyt$\rightarrow \pi^-+^3$He.
In these decays, the angular distribution of the decay products ($p$ and $^3$He) in the mother’s rest frame follows~\cite{STAR:2017ckg,Sun:2025oib}
\begin{equation}
\frac{dN}{\sin\theta^*d\theta^*}=\frac{1}{2}(1+\alpha_H\mathcal{P}_H\cos\theta^*),
\label{eq:weak}
\end{equation}
where $\alpha_H$ (subscript $H$ stands for $\Lambda$ and \hyt) is the decay  parameter, the polarization $\mathcal{P}_H=(N_{H,\frac{1}{2}}-N_{H,-\frac{1}{2}})/(N_{H,\frac{1}{2}}+N_{H,-\frac{1}{2}})$ characterizes the imbalance between the number of mother particle ($N_{H,\pm\frac{1}{2}}$) in different spin states, and $\theta^*$ denotes the angle between the daughter’s momentum and the spin direction of the mother particle. The decay parameter of \hyt~is linearly proportional to that of $\Lambda$, i.e., $\alpha_{^3_\Lambda\text{H}}\approx -\alpha_\Lambda/2.58$~\cite{Sun:2025oib}. These  decays are self-analyzing, as the momenta of the daughters carry direct information about the mother’s polarization.

\bmhead{Decoding proton spin polarization from $\Lambda$ and $^3_\Lambda$H polarizations} We now demonstrate how to unravel the proton spin polarization   from polarizations of hypertriton and  $\Lambda$ hyperon.
The spin polarization of \hyt~inherits those of nucleons and $\Lambda$ hyperons at their kinetic freeze-out. This becomes evident if one assumes that the polarizations of $\Lambda$ hyperons and nucleons are independent of their momenta, in which case the polarization of \hyt~from Eq.~\eqref{eq:polaCoal}  can be simplified as~\cite{Sun:2025oib}
\begin{eqnarray}
\label{eq:PolOneHalf1}
\mathcal{P}_{^3_\Lambda \text{H}}&\approx& \frac{2}{3}\mathcal{P}_n+\frac{2}{3}\mathcal{P}_p-\frac{1}{3}\mathcal{P}_\Lambda \approx \frac{1}{3} \left(4\mathcal{P}_p - \mathcal{P}_\Lambda\right),
\end{eqnarray}
where we have neglected higher-order terms (since polarizations are typically only a few percent) and assumed isospin symmetry, i.e., $\mathcal{P}_p \approx \mathcal{P}_n$.  The dominance of proton polarization in Eq.~(\ref{eq:PolOneHalf1}) explains why the $^3_\Lambda$H polarization changes only slightly from hadronization to  kinetic freeze-out in Fig.~\ref{fig:global_L} (c), in contrast to the larger change for $\Lambda$ polarization in Fig.~\ref{fig:global_L} (a). Inverting Eq.~\eqref{eq:PolOneHalf1} yields the expression for proton spin polarization given in Eq.~(\ref{eq:NucleonPolarization}).

Figure~\ref{fig:global_nucleon} (a) compares the model-calculated proton spin polarization (blue band) with the value of $(3\mathcal{P}_\text{\hyt} + \mathcal{P}_\Lambda)/4$ (orange band), obtained from the calculated polarizations of \hyt~and $\Lambda$ hyperons in Au+Au collisions. The two results are in excellent agreement, and their ratio ($(3\mathcal{P}_\text{\hyt} + \mathcal{P}_\Lambda)/4\mathcal{P}_p$), shown in the lower panel, remains consistent with unity within uncertainties. This confirms the validity of Eq.~(\ref{eq:NucleonPolarization}) as a reliable method to infer nucleon polarization from \hyt~and $\Lambda$ measurements. 

Considering that the momentum difference among nucleons and $\Lambda$ hyperons has to be very small when they form loosely-bound hypertritons via coalescence, Eq.~(\ref{eq:NucleonPolarization}) can be generalized to   account for momentum dependence, i.e.,
\begin{eqnarray}
\mathcal{P}_p (p_T) \approx \frac{1}{4}\left(3\mathcal{P}_{^3_\Lambda\mathrm{H}}(3p_T) + \mathcal{P}_\Lambda(p_T)\right),
\label{eq:NucleonPolarization2}
\end{eqnarray}
where $p_T$ denotes the transverse momentum.

Figure~\ref{fig:global_nucleon} (b) further shows the   $p_T$  dependence of    $\mathcal{P}_p(p_T)$ and $(3\mathcal{P}_\text{\hyt}(3p_T) + \mathcal{P}_\Lambda(p_T))/4$ in Au+Au collisions at $\sqrt{s_{NN}} = 7.7$ GeV.  Both quantities exhibit a decreasing trend with $p_T$ and remain in close agreement throughout. Their ratio, shown in the lower panel, stays near unity within uncertainties, further validating the effectiveness of Eq.~(\ref{eq:NucleonPolarization}) in capturing both the magnitude and momentum dependence of proton spin polarization. The same agreements hold for other collision  energies.
  
\bmhead{Conclusions} 
The spin information of protons and neutrons-the dominant baryonic constituents in nuclear collisions-is largely inaccessible  due to their stabilities.  
To enable such measurements, we have proposed a new route to unravel the proton spin polarization through independent measurements of the polarizations of $\Lambda$ hyperons and \hyt. To validate this method, we perform transport model calculations  incorporating spin-dependent scatterings and resonance feed-down effects during the hadronic phase of Au+Au collisions at $\sqrt{s_{NN}} = 7.7$–200 GeV.  We find that the proton polarization is related to the measurable $\Lambda$ and $^{3}_{\Lambda}\mathrm{H}$ polarizations through a simple and robust linear relation given by $\mathcal{P}_p\approx\left (3\mathcal{P}_\text{\hyt}+\mathcal{P}_\Lambda \right )/4$.

Accessing the proton spin polarization with our proposed method would opens a new window on the spin dynamics of up and down quarks, complementing existing insights from $\Lambda$ polarization that primarily probes strange quarks. It further enables studies of spin-dependent transport phenomena in hadronic matter~\cite{Liu:2023nkm,Xu:2025uwd}, provides constraints on the equation of state of spin-polarized nuclear matter~\cite{Tachibana:2025wey}, and helps resolving the proton spin puzzle in QCD~\cite{Ji:2020ena,Gross:2022hyw}.
Importantly, this method can be readily applied to ongoing and upcoming experiments at RHIC, the High Intensity heavy-ion Accelerator Facility (HIAF)~\cite{Zhou:2022pxl}, the Nuclotron-based Ion Collider Facility (NICA)~\cite{Kekelidze:2016hhw}, and the Facility for Antiproton and Ion Research (FAIR)~\cite{Durante:2019hzd}.

\backmatter

\bmhead{Acknowledgments}
The authors thank helpful discussions with Lie-Wen Chen,  Zuo-Tang Liang, Shu Lin, Xiao-Feng Luo, Guo-Liang Ma, Shi Pu, Yi-Feng Sun, Rui Wang, Jun Xu, Zhen Zhang, and Yong Zhao.
This work was supported in part by the National Key Research and Development Project of China under Grant No. 2024YFA1612500 and No. 2024YFA1610802;  the National Natural Science Foundation of China under contract No. 12422509, No. 12375121, No. 124B2102, No. 12147101, No. 12275054, No. 12025501, and No. 12061141008; the Natural Science Foundation of Shanghai under Grant No. 23JC1400200 and No. 23590780100; the Shanghai Pilot Program for Basic Research-Fudan University 21TQ1400100(22TQ006); the Guangdong Major Project of Basic and Applied Basic Research No. 2020B0301030008;  the STCSM under Grant No. 23590780100;  and the U.S. Department of Energy under Award No. DE-SC0015266.  The computations in this research were performed using the CFFF platform of Fudan University.  

\bmhead{Data availability}
All the data supporting the findings in this work are available within the manuscript and any additional data are available from the corresponding authors upon reasonable request.

\bmhead{Code availability}
Inquiries about the code in this work will be responded to by the corresponding authors.

\bmhead{Author contributions}
D. N. Liu performed the numerical simulations and prepared the figures. D. N. Liu and Y. P. Zheng conducted the analytical analysis. K. J. Sun and Y. G. Ma supervised the project and contributed to the interpretation of the results.   All authors participated in discussions and contributed to the preparation of the manuscript.

\bmhead{Competing interests}
The authors declare no competing interests.

\clearpage

\bmhead{\large{\bf Methods}} 
\bmhead{Spin polarization of hadrons at hadronization} 
We calculate the spin polarization of hadrons in Au+Au collisions at $\sqrt{s_{NN}} = 7.7$–200 GeV using the  AMPT model~\cite{Lin:2004en}, which has been extensively employed to describe a broad range of observables in relativistic heavy-ion collisions.  To describe the spin dynamics of hadrons during the hadronic matter expansion, we extend the AMPT model to include the spin degree of freedom of hadrons and further implement the spin-dependent scatterings and decays.

At hadronization, the spin polarization of hadrons is commonly determined by the thermal vorticity of the QGP under the assumption of local thermal equilibrium~\cite{Becattini:2007sr, Becattini:2013fla}. The leading-order spin polarization vector of a particle with non-zero spin $s$, mass $m$, four-momentum $p^\mu$, and produced at space-time point $x^\mu$, is given by~\cite{Becattini:2013vja, Becattini:2013fla, Fu:2020oxj}
\begin{equation}
     \mathcal{P}^\mu(x,p)=-\frac{s+1}{6m}(1-n_F)\epsilon^{\mu\nu\rho\sigma}p_\nu\varpi_{\rho\sigma}(x),
     \label{eq:pola}
\end{equation}
where $n_F = 1/[1 + \exp(\beta \cdot p \mp \mu/T)]$ is the Fermi-Dirac distribution function for particles $(-)$ and antiparticles $(+)$. In heavy-ion collisions, $n_F \approx 1$ due to the high temperature and low baryon and strangeness densities.  
In Eq.~\eqref{eq:pola}, the thermal vorticity tensor $\varpi_{\mu\nu}$ is defined as $\varpi_{\mu\nu}=-\frac{1}{2}(\partial_\mu\beta_\nu-\partial_\nu\beta_\mu)$ with $\beta^\mu=u^\mu/T$ being the four-velocity (satisfying the normalization condition $u^\mu u_\mu=1$) of the fluid divided by temperature. The temperature field and velocity field are obtained by solving the eigen equation $T^{\mu\nu} u_\mu = \epsilon u^\mu$, where $T^{\mu\nu}$ is the energy-momentum tensor and $\epsilon$ denotes the energy density of fluid cells in its local rest frame~\cite{Li:2017slc}.   The temperature field is estimated as $T=0.199~\text{GeV}\left(\frac{\epsilon/\gamma_q}{1+3N_f/4}\right)^{1/4}$~\cite{Lin:2014tya} with a fugacity factor $\gamma_q=\frac{1}{2}\left(\left(\frac{N_{+}}{N_{-}}\right)^{3/2} +\left(\frac{N_{-}}{N_{+}}\right)^{3/2}\right)$ where $N_+$ ($N_-$) represents the number of (anti)quarks in a cell summed over all events. Note that these fields are determined by averaging over a sufficiently large number (typically $10^5$) of events from AMPT and the spin polarization vector is calculated in the particle's rest frame with a Lorentz boost~\cite{Li:2017slc}.

\bmhead{Effects of hadronic  re-scatterings}
During the subsequent expansion of hadronic matter, the spin polarizations of nucleons and $\Lambda$ hyperons are modified by resonance feed-down and spin-dependent scatterings, with dominant contributions from $\Delta (1232)$ and $\Sigma^*(1385)$ resonances~\cite{Sung:2024vyc,Sung:2025pdj}.  For a two-body decay of the form $M \to D + X$, the spin polarization of the daughter particle ($\mathcal{P}_D$) is linearly proportional to that of the mother particle ($\mathcal{P}_M$), i.e., 
\begin{equation}
    {\cal{P}}_D={\cal C}_{M\rightarrow D}{\cal{P}}_{M},
\end{equation}
where $\mathcal{C}_{M\rightarrow D}$ is the spin transfer coefficient for the decay~\cite{Xia:2019fjf,Becattini:2016gvu,Karpenko:2016jyx}. For strong decays such as $\Delta \to N\pi$, $\Sigma^* \to \Lambda\pi$, and $N^* \to N\pi$, the spin transfer coefficient is $\mathcal{C}_{M\rightarrow D} = 1$.   In contrast, for the electromagnetic decay $\Sigma^0 \to \Lambda \gamma$, the spin transfer coefficient is negative, $\mathcal{C}_{\Sigma^0 \rightarrow \Lambda} = -\tfrac{1}{3}$~\cite{Xia:2019fjf}.  As a result, $\Lambda$ polarization is enhanced by feed-down from $\Sigma^*$ decays but suppressed by contributions from $\Sigma^0$.  Likewise, the feed-down from primary $\Delta$ resonances leads to an enhancement of nucleon polarization.  For weak decays $\Xi\to\Lambda\pi$ and $\Omega\to\Lambda K$, which are not included in the present study, their effects on  $\Lambda$ polarization are negligible because of the relative small number of $\Xi$ and $\Omega$~\cite{STAR:2019bjj,Sung:2025pdj}.

For the inverse process of resonance formation, such as $\pi + D \to R$, a similar linear relation holds, i.e.,
\begin{equation}
    {\cal{P}}_R={\cal C}_{D\rightarrow R}{\cal{P}}_D.
\label{eq:scatter}
\end{equation}
For  example, in the formation of spin-3/2 resonance $\Sigma^*$ via $\pi \Lambda \rightarrow \Sigma^*$, the spin transfer coefficient is $\mathcal{C}_{D \rightarrow R} = 5/9$~\cite{Xia:2019fjf, Yang:2017sdk}, corresponding to a $\Lambda$ spin flip to non-flip  ratio of $2/7$ in the $s$ channel~\cite{Sung:2024vyc}. The same value, $\mathcal{C}_{D \rightarrow R} = 5/9$, applies to $\Delta$ formation via $\pi N \rightarrow \Delta$.
In contrast, for the formation of a spin-1/2 resonance such as $N^*$ via $\pi N \rightarrow N^*$, the spin transfer coefficient is $\mathcal{C}_{D \rightarrow R} = 1$. 
Consequently,   quasi-elastic   scatterings such as $\pi \Lambda \rightarrow \Sigma^*  \rightarrow \pi \Lambda$ and $\pi N \rightarrow \Delta  \rightarrow \pi N$ reduce the spin polarization of primary $\Lambda$ hyperons and nucleons to approximately $5/9$ of their initial values at hadronization. 
As shown in Fig.~\ref{fig:global_L} (a), the combined effects of resonance formation and decay suppress the $\Lambda$ polarization by about 20\%, consistent with results obtained from solving rate equations with a parametrized fireball expansion~\cite{Sung:2024vyc,Sung:2025pdj}.

\bmhead{Hypertriton production within a spin-dependent coalescence model}   In the covariant coalescence model, the invariant momentum distribution of hypertritons formed from the coalescence of polarized protons, neutrons, and $\Lambda$ hyperons is given by~\cite{Sun:2025oib,Yang:2017sdk}
\begin{eqnarray}
\label{eq:Coal} 
E\frac{\text{d}^3N_{^3_\Lambda \text{H},\pm\frac{1}{2}}}{\text{d}{\bf P}^3} &=& E\int \prod_{i=n,p,\Lambda}p_i^\mu\text{d}^3\sigma_{\mu}\frac{\text{d}^3p_i}{E_i}\bar{f}_i({\bf x}_i,{\bf p}_i)  \notag \\
&\times&\left(\frac{2}{3}w_{n,\pm\frac{1}{2}}w_{p,\pm\frac{1}{2}}w_{\Lambda,\mp\frac{1}{2}}\right.  \notag \\
&&+\frac{1}{6}w_{n,\pm\frac{1}{2}}w_{p,\mp\frac{1}{2}}w_{\Lambda,\pm\frac{1}{2}} \notag \\
&&+\left.\frac{1}{6}w_{n,\mp\frac{1}{2}}w_{p,\pm\frac{1}{2}}w_{\Lambda,\pm\frac{1}{2}}\right) \notag \\
&\times& W_{^3_\Lambda \text{H}} \delta^{(3)}({\bf P}-\sum_i{\bf p}_i),
\end{eqnarray}
where $p^\mu$ is the four-momentum, $d\sigma_{\mu}$ represents the normal vector of the hypersurface of freeze-out particles, $\bar{f}({\bf x}_i,{\bf p}_i)$ denotes their spin-averaged phase-space distributions with the coordinate ${\bf x}_i$ and momentum ${\bf p}_i$ in the frame of the emission source, and $W_{\text{\hyt}}$ is the  Wigner function of \hyt, which we take to have a Gaussian form, $W_{^3_\Lambda\text{H}} = 8^2 \exp \left[ - \frac{\rho^2}{\sigma_{\rho}^2} - \frac{\lambda^2}{\sigma_{\lambda}^2} - p_{\rho}^2 \sigma_{\rho}^2 - p_{\lambda}^2 \sigma_{\lambda}^2 \right]$ with $\rho$, $\lambda$, $p_\rho$, and $p_\lambda$ denoting the relative distance in coordinate and momentum spaces~\cite{Liu:2024ygk}.   The parameters used in the Wigner function are $\sigma_\rho=2.26$ fm to reproduce the empirical deuteron root-mean-squared radius of $\sqrt{\left < r_d^2 \right>}=1.96$ fm~\cite{Scheibl:1998tk} and $\sigma_\lambda=6.52$ fm, which corresponds to the world average $\Lambda$ separation energy of  $B_\Lambda=0.164$ MeV in  \hyt. Using the phase-space information of  nucleons and $\Lambda$ hyperons at kinetic freeze-out (i.e., their positions and momenta at the last interactions) from the extended AMPT model, the integrated yield of hypertriton from Eq.~(\ref{eq:Coal}) is simplified to Eq.~(\ref{eq:polaCoal}). 


\end{document}